\documentclass[useAMS,usenatbib]{mnras}
\usepackage{natbib}
\usepackage{mathtools}
\usepackage{amsmath}
\usepackage{amssymb}

\title[The Sparkler's GCs]{The Ages and Metallicities of the Globular Clusters in the Sparkler}

\author[Adamo et al.]{
Angela~Adamo,$^{1}$
Christopher~Usher,$^{1}$
Joel Pfeffer,$^{2}$ and
Ad\'ela\"ide~Claeyssens$^{1}$\\
$^{1}$The Oskar Klein Centre, Department of Astronomy, Stockholm University, AlbaNova, SE-106 91 Stockholm, Sweden \\
$^{2}$Centre for Astrophysics \& Supercomputing, Swinburne University, Hawthorn, VIC 3122, Australia 
}

\date{Accepted 2022 June 20. Received 2022 June 20; in original form 2023 March 21}

\pubyear{2023}

\begin{document}
\label{firstpage}
\pagerange{\pageref{firstpage}--\pageref{lastpage}}
\maketitle

\begin{abstract}
JWST observations of the strongly lensed galaxy The Sparkler have revealed a population of gravitationally bound globular cluster (GC) candidates. Different analyses have resulted in broadly similar ages but significantly different metallicities, questioning the assembly history that has led to the formation of such a population. In this letter, we re-analyse the two sets of photometry available in the literature with the code  \textsc{mcmame} especially tailored to fit physical properties of GCs. We find the ages and metallicities from both datasets are consistent within 1$\sigma$ uncertainties. A significant group of GCs is consistent with being old and metal poor ([Fe/H] $\sim -1.7$).  For this group, the ages do not converge, hence, we conclude that they are definitively older than 1 Gyr and can be as old as the age of the Universe. The remaining GCs have younger ages and a metallicity spread. The ages and metallicities distribution of GCs in the Sparkler are consistent with those observed in Local Group’s galaxies at similar lookback times. Comparing with predictions from E-MOSAICS simulations we confirm that the Sparkler GC population traces the self-enrichment history of a galaxy which might become a few times $10^9$ M$_{\odot}$ massive system at redshift $z = 0$.
\end{abstract}


\vspace{-1cm}
\section{Introduction}

The first JWST observations of a cosmological lensed field \citep[SMACS0723,][]{ERO} have enabled the detection of globular cluster (GC) candidates in a galaxy, the Sparkler, close to the peak of the cosmic formation history, e.g., redshift $z = 1.38$ \citep[][hereafter M22]{Mowla2022}. 
Sizes and intrinsic physical properties are consistent with these stellar systems to be gravitationally bound \citep[][hereafter C23]{claeyssens2023}. GCs have long been considered remnants of the past assembly history of their host galaxies \citep{brodie2006}. The combination of JWST and magnification by gravitational telescopes has been reported as a powerful tool to enable the detection of GCs at high redshift \citep{renzini17, Forbes2018}. Indeed, initial JWST studies of lensed galaxies have reported potentially gravitational bound young ($\sim 10$ Myr) proto-GCs up to $z \sim 6$, i.e., at the edge of reionisation \citep{Vanzella2022SB}. Simulations predict the formation of proto-GCs beginning in the reionisation era \citep{reinacampos2022, garcia2022}, although their survival rates depend on tidal disruption \citep{reinacampos2022, meng2022}. \\
\indent The discovery of the GC population in the Sparkler has raised great interest in the community because of the reported redshift formation of these systems ($z\sim9$) as well the implications for the assembly history of its host galaxy. Two independent analyses have been conducted so far. \\
\indent In the first study, M22 extracted with fixed aperture photometry the integrated spectral energy distribution (SED) of the 9 clusters surrounding the system. They performed a nonparametric star formation history (SFH) SED fitting analysis with the code \texttt{DENSE BASIS}. Their recovered solutions suggest that 5 of the 9 objects have ages of $\sim 3.9-4.1$ Gyr at redshift $z = 1.38$, consistent with a redshift of formation between 7 and 11. Surprisingly the metallicity of these GCs are all significantly high, with abundances between 20 and 75 \% Solar ($-0.7 \lesssim$ [Fe/H] $\lesssim -0.1$).  The remaining 
4 systems have ages below 300 Myr and display a similar metallicity range. \\
\begin{figure*}
\centering
\includegraphics[width=13.5cm]{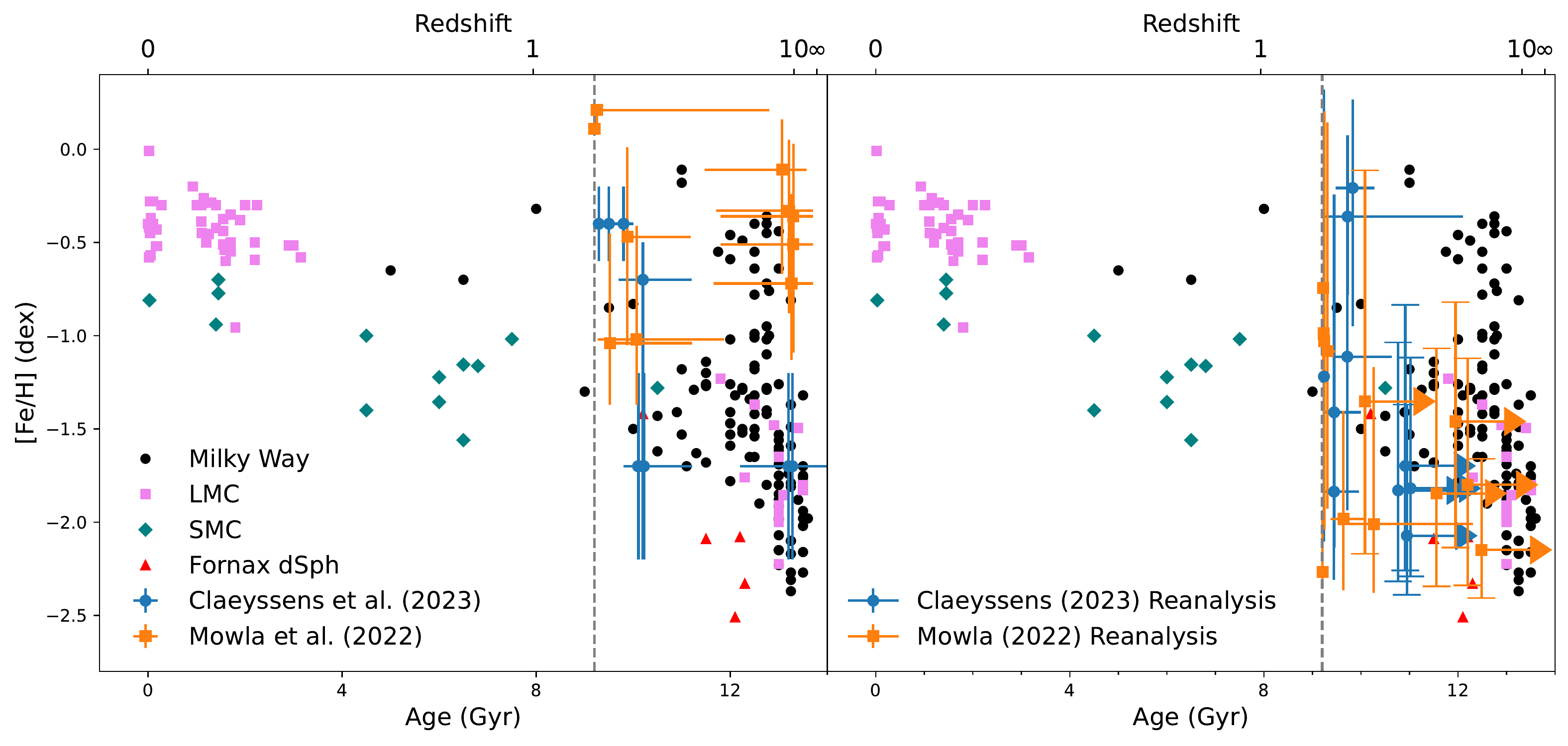}
\caption{The ages and metallicity of GCs in the Sparkler compared to GCs in the Milky Way (black circles), the Large Magellanic Cloud (violet squares), the Small Magellanic Cloud (teal diamonds) and the Fornax dwarf (red triangles).
These ages and metallicities are drawn from the sample of \citet{2019MNRAS.482.1275U}. 
In the left panel, we show the ages and metallicities from \citet[blue circles][C23]{claeyssens2023} and \citet[orange squares][M22]{Mowla2022}.
In the right we show the medians and 68 \% confidence intervals of the posteriors determined in this work.  Lower limits are used for the clusters where the ages remain unconstrained (see text). The Sparkler redshift is indicated by the vertical dashed gray line.}
\label{fig:combined}
\end{figure*}
\indent In the second analysis, C23 use a Gaussian fitting approach to simultaneously determine the size and fluxes of 10 clusters surrounding the Sparkler. The SED fitting analysis has been performed using single stellar population models \citep[Yggdrasil,][]{Zackrisson2011} with 4 different metallicity steps ($Z=0.0004$ to 0.02, the latter considered Solar). Two of the 8 GC candidates (in common between the two works) have ages of 4 Gyr (thus, in agreement with the M22' analysis) but with significantly lower metallicity ([Fe/H] $\sim -1.7$). The remaining systems have ages between 0.1 and 1 Gyr and significant spread in metallicity ($-1.7 \leq$ [Fe/H] $\leq -0.4$). The left panel of Figure~\ref{fig:combined} illustrates the age and metallicities of the Sparkler clusters as derived by M22 and C23 and their degree of disagreement.\\
\indent By exclusively using the M22 results, \citet{forbes2023} argue that the Sparkler is the progenitor of a Milky Way (MW) like galaxy but with significant differences. As visible in the left panel of Figure~\ref{fig:combined}, the Sparkler would be missing the very old and metal-poor GCs typically observed in Local Group galaxies. The lack of the latter population is in disagreement with models of galaxy assembly that take into account self-enrichment \citep[e.g.,][]{kruijssen2019b, Horta2021}, and would point toward an unusually fast enrichment and rapidly assembly for the Sparkler.\\
\indent In this letter, we revisit the \citet{forbes2023} interpretations of the the Sparkler GC population and its galaxy assembly history using a different approach. We re-analyse the GC SEDs independently published by M22 and C23 using the latest state-of-the-art single stellar population libraries widely used to study integrated light of GCs in the local Universe. Using posterior-distributions, we derive constraints on the age, metallicity, extinction, mass of the GCs in the two datasets. We establish the level of agreement between the two sets of photometric analyses and we compare the recovered ages and metallicities of the Sparkler with the age-mass relation (AMR) of observed GC populations observed in the Local Group (MW, LMC, SMC, Fornax), as well as predicted from cosmological simulations that analytically develop formation and evolution of star clusters \citep[E-MOSAICS,][]{pfeffer18}. We assume the \citet{2020A&A...641A...6P} $\Lambda$CMD cosmology parameters .

\begin{figure}
    \centering
    \includegraphics[width=240pt]{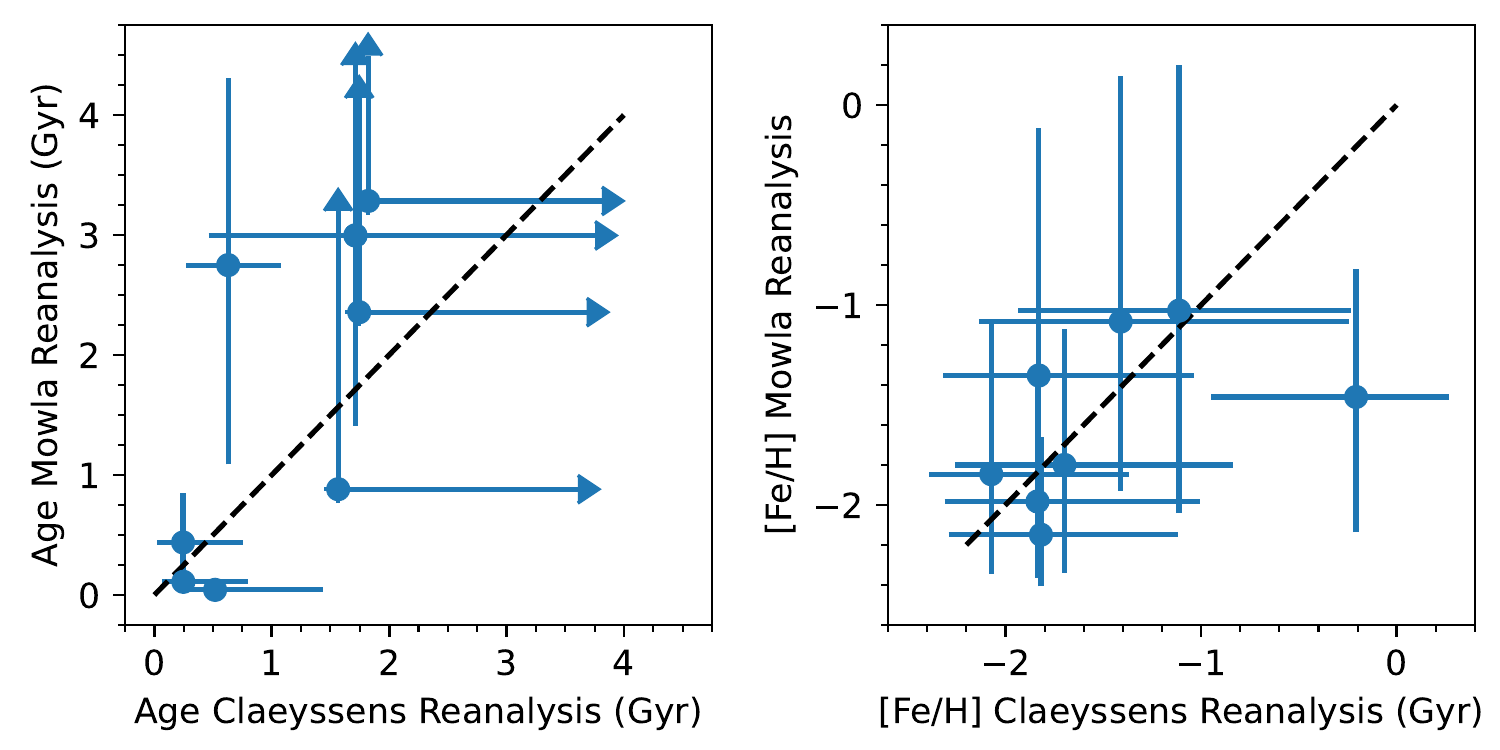}
    \caption{Comparison of the GC ages (left) and metallicities (right) we derive from the C23 and M22 photometry, respectively.
    The error bars give the medians and the 16 to 84 \% percentiles of the posteriors. For the old GCs we reported their ages as lower limits due to the lack of convergence in the marginalised distributions. The dashed line is the one-to-one relation.}
    \label{fig:mowla_comparison}
\end{figure}

\vspace{-0.5cm}
\section{Analysis}

We first use the published JWST NIRCam photometry of the GCs surrounding the Sparkler in the bands F090W, F150W, F200W, F277W, F356W and F444W by C23.

\begin{figure*}
    \centering
    \includegraphics[width=17.5cm]{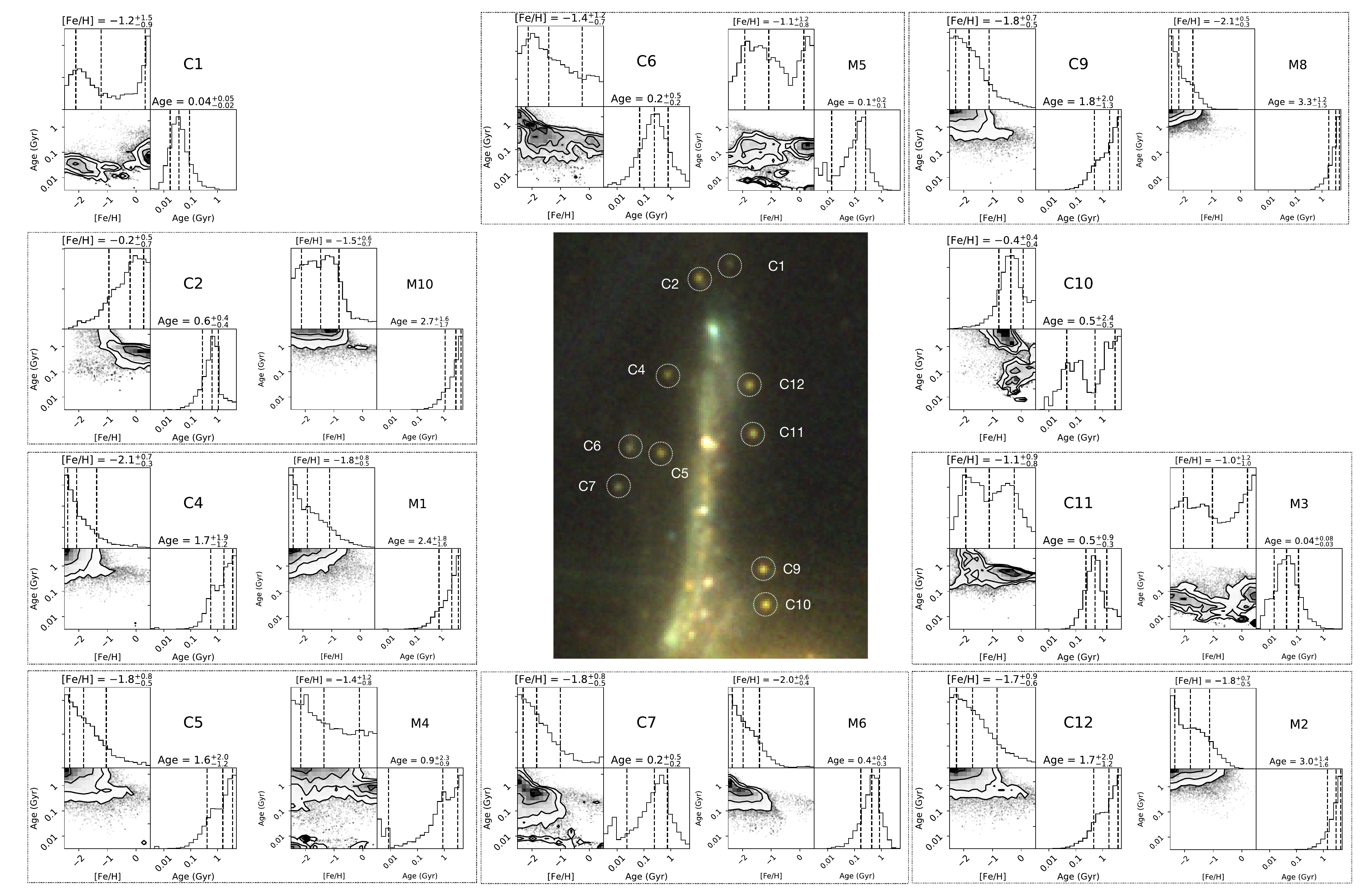}
    \caption{Age-metallicity posterior distributions and SED for the Sparkler's GCs for the C23 and M22 reanalyses. The systems are matched when overlapping in both datasets and named according to the two nomenclatures.
    In posterior distributions, the vertical dashed lines show the 16, 50 and 84 percentiles. 
    In the centre we show a NIRCam F090W, F150W and F200W (roughly restframe $uri$) image from C23 with the GCs analysed in this work highlighted.}
    \label{fig:Sparkler_GC}
\end{figure*}

The Monte Carlo Markov Chain code \textsc{mcmame} \citep[Usher et al. submitted]{2019MNRAS.490..491U} is used to sample the age, metallicity, mass and reddening posteriors of a grid of stellar population models subject to the constraints provided by the JWST NIRCam photometry.
 We use \textsc{fsps} \citep{2009ApJ...699..486C, 2010ApJ...712..833C}  to calculate a model grid at $z = 1.378$ with the MIST isochrones \citep{2016ApJ...823..102C},  the MILES spectral library \citep{2006MNRAS.371..703S} and the \citet{1989ApJ...345..245C} extinction curve.
We run \textsc{mcmame} with a uniform prior on metallicity in the range $-2.5 < $ [Fe/H] $ < 0.5$, on age in the range between 1 Myr and 5 Gyr (lookback time of the universe at that redshift), positive extinction ($A_{V} > 0$) and on $\log$ mass.
Before fitting, we correct the photometry for the Milky Way's foreground extinction reported in table 1 of C23.

We plot the recovered median and 68 \% intervals of the age and metallicity posteriors in the right hand panel of Figure \ref{fig:combined} along with those of the GCs of MW and its satellite galaxies. For the oldest clusters where we do not get a convergence in the age estimation (see Figure~\ref{fig:Sparkler_GC}), we plot the median as a lower limit, to reflect that the ages of these systems remain unconstrained. Our ages and metallicities are qualitatively in agreement with those found by C23 (left panel). We recover  a group of GCs with older ages and low metallicities and a group of GCs with young ages and a range of metallicities. These recovered values are also in agreement with the observed positions the clusters in the color-color diagram showed in Figure 13 of C23.\\
\indent Secondly, we re-analysed the M22 photometry.
We plot a comparison of the medians of the age and metallicity posteriors we obtained from the C23 and the M22 photometry in Figure~\ref{fig:mowla_comparison}.
In Table~\ref{table}, we report the 16, 50, 84 \% values for each fitted parameter from the two sets of photometry. Overall, by applying the same methodology to fit the observed GC SEDs, we derive solutions consistent within 1$\sigma$ from each others in the majority of the cases. We conclude that both methods to perform photometry are consistent.
 We speculate that the differences between the physical properties derive for the GC population of the Sparkler by M22 and C23 arise from both the different methods used to analyse the SEDs  and underestimations of the resulting physical parameter uncertainties. The latter could be largely reduced with more precise photometry and deeper data covering the 0.3 to 2 $\mu$m restframe.
As for the SED analysis, we notice that both our reanalysis and that of M22 utilise the same stellar population models (\textsc{fsps}, \citealt{2009ApJ...699..486C, 2010ApJ...712..833C}) although M22 use models calculated with the BaSeL stellar library \citep{2002A&A...381..524W} and the Padova isochrones \citep{2008A&A...482..883M} rather than the MILES library \citep{2006MNRAS.371..703S} and MIST isochones \citep{2016ApJ...823..102C} used in this work. Differences between the two stellar libraries might be the source of the disagreement, although we notice that Yggdrasil models used by C23 are based on Padova stellar libraries \citep{Zackrisson2011}. \\
\begin{table*}
    \centering
\begin{tabular}{cccccccccc}
C23 & M22 & \multicolumn{2}{c}{C23} & \multicolumn{2}{c}{M22} & \multicolumn{2}{c}{C23 Reanalysis} & \multicolumn{2}{c}{M22 Reanalysis} \\ 
 ID & ID & Age  [Gyr] & [Fe/H] & Age  [Gyr] & [Fe/H] & Age [Gyr] & [Fe/H] & Age  [Gyr] & [Fe/H]  \\ \hline 
 C1 & --  & $0.60_{-0.50}^{+0.20}$ & $-0.4_{-0.2}^{+0.2}$ & -- & -- & $0.04_{-0.02}^{+0.05}$ & $-1.2_{-0.9}^{+1.5}$ & -- & -- \\ 
C2 & M10  & $4.00_{-1.00}^{+1.00}$ & $-1.7_{-0.5}^{+0.5}$ & $4.10_{-1.50}^{+0.41}$ & $-0.4_{-0.5}^{+0.4}$ & $0.63_{-0.35}^{+0.45}$ & $-0.2_{-0.7}^{+0.5}$ & $>2.75_{-1.66}^{+1.56}$ & $-1.5_{-0.7}^{+0.6}$ \\ 
C4 & M1  & $4.08_{-1.00}^{+1.00}$ & $-1.7_{-0.5}^{+0.5}$ & $4.10_{-1.50}^{+0.41}$ & $-0.5_{-0.6}^{+0.5}$ & $>1.74_{-1.19}^{+1.94}$ & $-2.1_{-0.3}^{+0.7}$ & $>2.36_{-1.57}^{+1.78}$ & $-1.8_{-0.5}^{+0.8}$ \\ 
C5 & M4 & $1.00_{-0.20}^{+1.00}$ & $-1.7_{-0.5}^{+0.5}$ & $3.87_{-1.60}^{+0.50}$ & $-0.1_{-0.6}^{+0.3}$ & $>1.57_{-1.16}^{+2.04}$ & $-1.8_{-0.5}^{+0.8}$ & $>0.88_{-0.87}^{+2.32}$ & $-1.4_{-0.8}^{+1.2}$ \\ 
C6 & M5 & $0.30_{-0.10}^{+0.10}$ & $-0.4_{-0.2}^{+0.2}$ & $0.32_{-0.00}^{+1.70}$ & $-1.0_{-0.3}^{+0.6}$ & $0.25_{-0.18}^{+0.55}$ & $-1.4_{-0.7}^{+1.2}$ & $0.11_{-0.10}^{+0.15}$ & $-1.1_{-0.8}^{+1.2}$ \\ 
C7 & M6 & $0.09_{-0.05}^{+0.11}$ & $-0.4_{-0.2}^{+0.2}$ & $0.87_{-0.80}^{+1.80}$ & $-1.0_{-0.4}^{+0.6}$ & $0.24_{-0.22}^{+0.51}$ & $-1.8_{-0.5}^{+0.8}$ & $0.44_{-0.27}^{+0.41}$ & $-2.0_{-0.4}^{+0.6}$ \\ 
C9 & M8 & $0.91_{-0.20}^{+0.10}$ & $-1.7_{-0.5}^{+0.5}$ & $4.06_{-1.60}^{+0.45}$ & $-0.7_{-0.4}^{+0.5}$ & $>1.82_{-1.33}^{+1.99}$ & $-1.8_{-0.5}^{+0.7}$ & $>3.28_{-1.51}^{+1.21}$ & $-2.1_{-0.3}^{+0.5}$ \\ 
C10 & -- & $1.00_{-0.50}^{+1.00}$ & $-0.7_{-0.5}^{+0.2}$ & -- & -- & $0.52_{-0.48}^{+2.38}$ & $-0.4_{-0.4}^{+0.4}$ & -- & -- \\ 
C11 & M3  & $1.02_{-0.10}^{+1.00}$ & $-1.7_{-0.5}^{+0.5}$ & $0.68_{-0.00}^{+1.30}$ & $-0.5_{-0.6}^{+0.5}$ & $0.52_{-0.28}^{+0.92}$ & $-1.1_{-0.8}^{+0.9}$ & $0.04_{-0.03}^{+0.08}$ & $-1.0_{-1.0}^{+1.2}$ \\ 
C12 & M2 & $0.90_{-0.30}^{+0.10}$ & $-1.7_{-0.5}^{+0.5}$ & $4.01_{-1.50}^{+0.50}$ & $-0.3_{-0.6}^{+0.4}$ & $>1.71_{-1.25}^{+2.04}$ & $-1.7_{-0.6}^{+0.9}$ & $>3.00_{-1.59}^{+1.42}$ & $-1.8_{-0.5}^{+0.7}$ \\ 
\end{tabular}
    \caption{Ages and metallicities from \citet[][M22]{Mowla2022} \citet[][C23]{claeyssens2023} and this work. The ages of the old GCs are reported as lower limits due to the lack of convergence in the marginalised distributions (Figure~\ref{fig:Sparkler_GC}), reflecting that these systems can be as old as the age of the universe.
    For C23 we give the age of the metallicity with the lowest reduced $\chi^{2}$.
    For M22 and our measurements we give the medians and the 16 to 84 \% of the posteriors.}
    \label{table}
\end{table*}
\indent In Figure~\ref{fig:Sparkler_GC} we identify the position of each GC using in the 3-color image of the Sparkler applying the naming convention of C23.
For each cluster,  we show the corner plots of the recovered age and metallicity for both C23 and M22 reanalyses when available, matching their respective identities. 
 The corner plots show significant degeneracies and non-Gaussian posteriors  in both datasets. The metallicities seem to be  constrained in the majority of the cases. On the other hand, the ages do not converge for the older GCs, where we can only conclude that they are definitively older than 1 Gyr and can be as old as the age of the Universe. For younger clusters we see that age constrains are significantly tighter, while the metallicity remains in some cases less constrained. The large uncertainties associated to the data in the right side of Figure~\ref{fig:combined} simply reflect the level of convergence in the determination of the two quantities in both datasets.
\vspace{-0.5cm}
\section{Discussion \& Conclusion}
\begin{figure*}
    \centering
    \includegraphics[width=7.55cm]{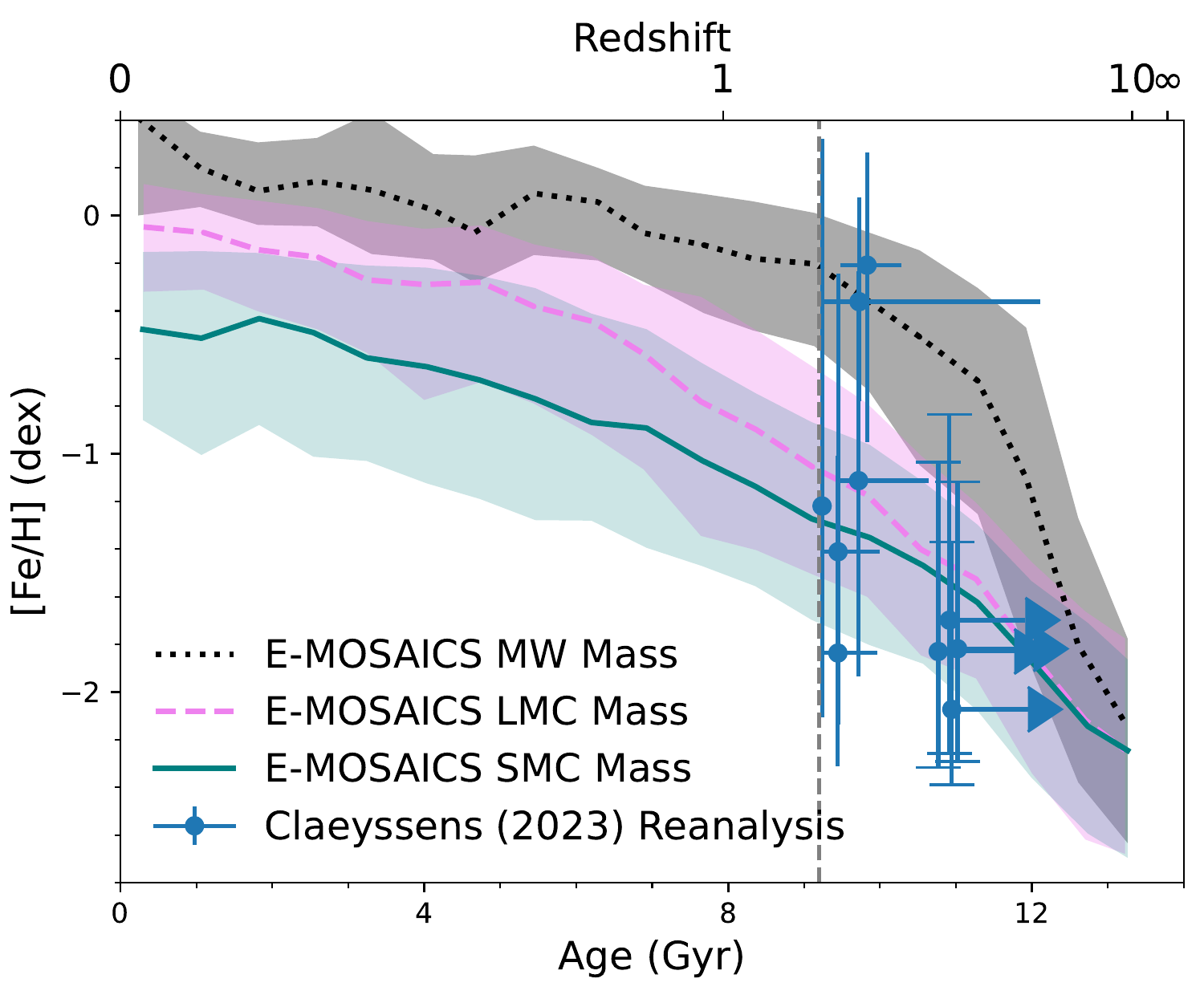}
    \includegraphics[width=7.8cm]{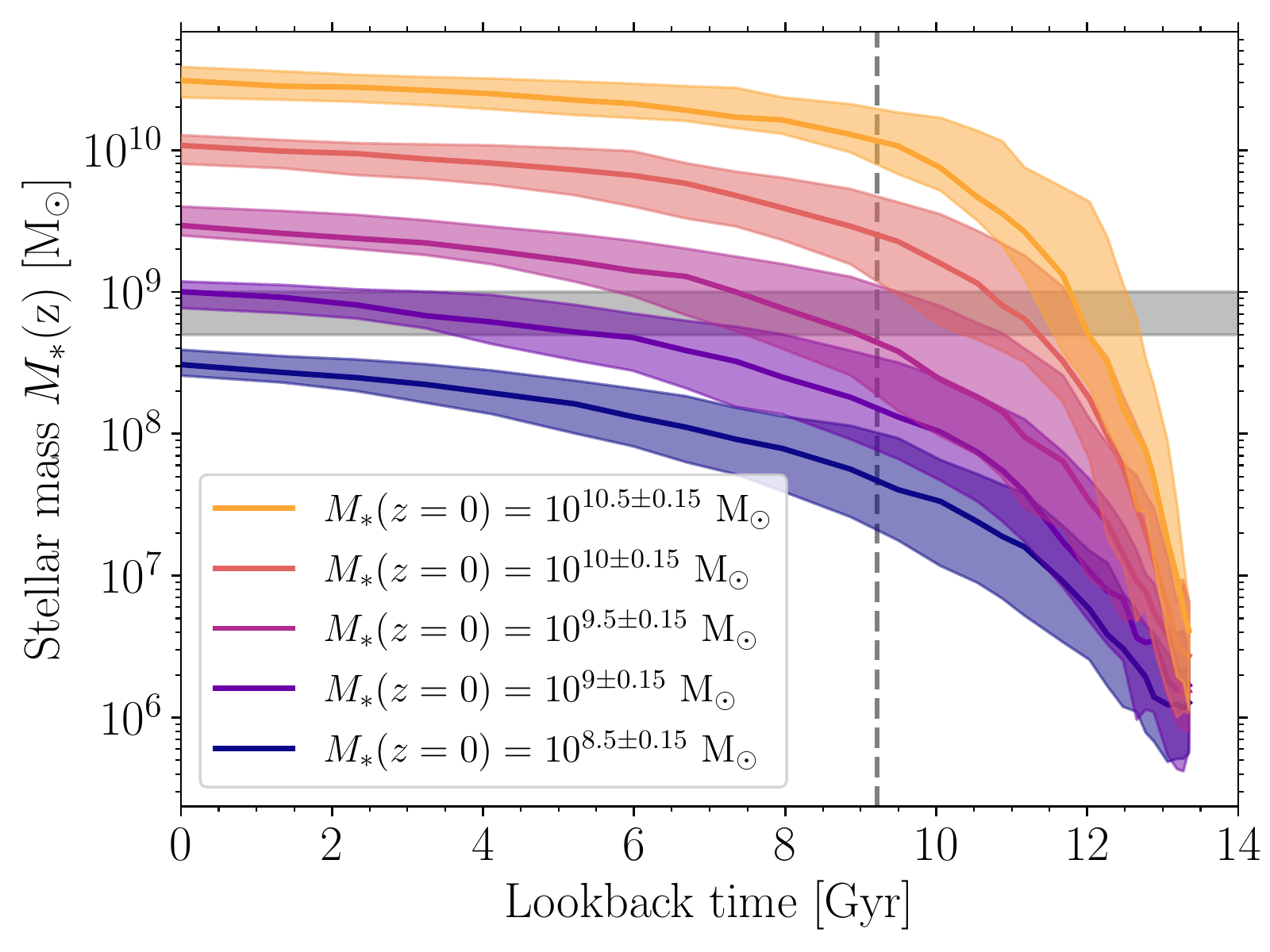}
    \caption{Left: Comparisons of our median age and metallicity posteriors (blue error bars) with the predictions of the E-MOSAICS simulations \citep{Horta2021}.  The M22 points are omitted for clarity.
    The lines show the median GC metallicities at a given age while the shaded regions show the 16 to 84 \% range for galaxies in MW-mass haloes (dotted black line, $0.7 \times 10^{12}\  \text{M}_{\odot}  < M_{200}< 3 \times 10^{12}\ \text{M}_{\odot}$), LMC mass galaxies (dashed green line, $10^{9}\  \text{M}_{\odot}  < M_{*}< 2 \times 10^{9}\ \text{M}_{\odot}$) and SMC mass galaxies (solid red line, $3 \times 10^{8}\  \text{M}_{\odot}  < M_{*}< 7 \times 10^{8}\ \text{M}_{\odot}$).
    The Sparkler's GCs are consistent with the age-metallcity relation of a galaxy between the mass of MW and the LMC today. Right: Median growth of galaxies extracted from the EAGLE simulation \citep{schaye15, crain15}. The final total stellar masses at redshift $z = 0$ are reported in the caption. The dashed line and the grey band shows the position and mass range (respectively) of the Sparkler at $z = 1.38$.}
    \label{fig:e_mosaics}
\end{figure*}

The Sparkler galaxy, magnified by the gravitational potential of the galaxy cluster SMACS0723, is a rather low mass galaxy at redshift $z = 1.38$. M22 report a total stellar mass for the host of $\log(M_*/M_{\odot}) \sim 9.7$, which corrected for lensing effect implies an intrinsic mass between $5\times10^8$ and $1\times10^9$ M$_{\odot}$ (see C23 for discussion of the different lensing model predictions).\\
\indent Taking advantage of the published photometry for the GC candidates by M22 and C23, 
 we reanalyse the two datasets with the same framework. In both datasets, we find a group of old and metal-poor GCs. Their ages do not converge, leading us to conclude that they are definitively older than 1 Gyr and can be as old as the age of the Universe. The other group coincides with younger clusters. The latter age constrains are significantly tighter, while uncertainties in the metallicity remain large. A recently accepted JWST cycle 2 program (GO 2969) will obtain NIRSpec spectroscopy of the 0.4 to 2 $\mu$m restframe, enabling to narrow down ages and metallicities for the cluster population.\\
\indent In general, we conclude that the Sparkler harbours metal-poor GCs that have ages overlapping with the metal-poor sequence of Local Group galaxies.  The younger GCs have formed with higher metallicities. The spread in metallicities in the younger GCs could support the scenario of a rapid accretion/merger with a satellite galaxy. The positions of the GCs around the Sparkler clearly suggest that the galaxy has undergone rapid dynamical evolution which has caused the GCs to remain located around the galaxy in a similar fashion to GCs observed in galaxies in the local Universe. We cannot conclude whether the metal-poor(rich) GCs have been formed \emph{in-situ} or \emph{ex-situ}, but the age of the younger GCs suggest that the interaction that led to the ejection of the younger population has happened less than 1 Gyr ago, i.e., at the peak of cosmic noon.\\
\indent In the left panel of Figure \ref{fig:e_mosaics}, we compare the derived ages and metallicities to the GC age-metallicity relations of different mass galaxies from the E-MOSAICS suite of simulations \citep{pfeffer18, kruijssen2019a}.
E-MOSAICS combines the EAGLE \citep{schaye15, crain15} hydrodynamic model of cosmological galaxy formation with subgrid models for the formation and evolution of star clusters \citep{kruijssen11, pfeffer18}. 
The GC age-metallicity relations for MW, LMC and SMC-mass galaxies were presented in \citet{Horta2021}, where the relation for MW-mass galaxies are derived from 25 zoom-in simulations \citep{kruijssen2019a} and the relations for LMC and SMC-mass galaxies are derived from a $34.4^{3}$ comoving-Mpc$^{3}$ volume \citep[first presented in][]{bastian20}.
Overall, we see that the Sparkler GCs follow approximately the age--metallicity relation expected by GCs forming in host galaxies that are growing in the E-MOSAICS simulations to became analogs of galaxies observed in the Local Group. On the right panel, we plot the median growth predictions for a range of different galaxy stellar masses at redshift $z = 0$ from the EAGLE \textsc{Recal-L25N752} simulation \citep{schaye15, crain15}. The position of the Sparkler is highlighted in grey, suggesting that the Sparkler might possibly become a few times $10^9$ M$_{\odot}$ galaxy at $z = 0$, similar to M33 \citep[$M_\ast \approx3 \times 10^9\  \text{M}_{\odot}$,][]{McConnachie_12}, e.g., between the LMC \citep[$M_\ast \approx1.5 \times 10^9\ \text{M}_{\odot}$,][]{McConnachie_12}  and the MW \citep[$M_\ast \approx 5 \times 10^{10}\ \text{M}_{\odot}$][]{Bland-Hawthorn_and_Gerhard_16} galaxies. These conclusions differ from those reached by \citet{forbes2023} whom suggested, using only the M22 results, that the Sparkler could be a MW progenitor without the metal-poor old GC population.\\
\indent GC populations have long been established as remnants of past events in the assembly history of galaxies \citep[e.g.][]{brodie2006}. In particular, the GC age--metallicity relation has early on been recognised as a powerful tool to reconstruct the assembly history of our own galaxy \citep{forbes2010}, as well as in the Local Group \citep[e.g.][for a recent review]{Forbes2018} and beyond \citep{2019MNRAS.490..491U}. Decoding observed GC AMR, however, it is not straight forward, because it requires to trace back GCs formed \emph{in-situ} vs. those accreated from other satellite galaxies, thus, formed \emph{ex-situ}. Cosmological simulations of Milky Way-like galaxies (as done in E-MOSAICS) which analytically develop the formation and evolution of star clusters has shown the AMR to vary with galaxy assembly history at fixed stellar mass \citep[e.g][]{kruijssen2019a}. Both observations and simulations agree that the Milky Way has a steep AMR for the \emph{in-situ} GCs, suggesting a rapid assembly and metal enrichment of the host, while younger metal-poor GC sequences are the results of accretions from lower-mass satellites \citep{kruijssen2019a}. On the other hand, GCs forming in galaxies like the Magellanic clouds result in shallower AMR \citep{Horta2021}. In the case of the Sparkler, the comparison with E-MOSAICS derived GC AMR (left plot), and average galaxy growth (right plot of Figure~\ref{fig:e_mosaics}) suggest that the Sparkler GC AMR is shallower than the one expected and observed for a Milky Way type galaxy. The large metallicity spread at younger ages could be evidence of a recent merger that can also explain why the GCs are located around the main body of the galaxy, in a configuration that is also observed in redshift $z = 0$ galaxies.    \\
\indent To conclude, the combination of gravitational lensing and JWST sensitivity and resolution has enabled the unprecedented detection of GCs surrounding a star-forming galaxy at redshift about $z = 1.4$. We show that the Sparkler GCs fit well-within the derived assembly history of galaxies in our Local Group from both numerical simulations as well as extrapolated from the observed GC populations. However, this results cannot be generally applied - the Sparkler is a single galaxy with a unique assembly history. With the increasing effort to survey more lens regions with JWST we expect a significant increase in the number of detected proto-GC as well as evolved GC populations. As GCs trace the growth of their host galaxies, we expect to reveal a broader spectrum of galaxy assembly histories than accessible in the Local Group.

\vspace{-0.5cm}
\section*{Acknowledgements}
The authors thank the referee for a constructive report. AA, AC and CU acknowledge support from the Swedish Research Council, Vetenskapsr{\aa}det (2021-05559, 2016-05199).
JP is supported by the Australian government through the Australian Research Council's Discovery Projects funding scheme (DP220101863). The authors thank the E-MOSAICS team for kindly sharing access to their simulations.
This work made use of \textsc{numpy} \citep{numpy}, \textsc{scipy} \citep{scipy}, \textsc{matplotlib} \citep{matplotlib}, \textsc{corner} \citep{corner} and \textsc{astropy} \citep{2013A&A...558A..33A}.

\vspace{-0.5cm}
\section*{Data Availability}
Data used in the analysis are publicly available.  

\vspace{-0.5cm}
\bibliographystyle{mnras}
\bibliography{bib}{}

\label{lastpage}

\end{document}